\documentclass[prl,aps,A4paper,singlecolumn,superscriptaddress,preprintnumbers,nopacs,tightenlines,floatfix,amsmath,amssymb]{revtex4}

\usepackage{bm}

\usepackage{amsmath}
\usepackage{graphicx}
\usepackage{amssymb}
\usepackage{mathrsfs}
\usepackage{bbm}

\newsavebox{\marginbox}

\newenvironment{comment}{%
  \begin{lrbox}{\marginbox}
  \begin{minipage}{\marginparwidth}
    \footnotesize
}{
  \end{minipage}
  \end{lrbox}
  \marginpar{\usebox{\marginbox}}
}

\makeatother

\usepackage{lipsum}

\newcommand{\be}{\begin{eqnarray}}
\newcommand{\ee}{\end{eqnarray}}

\begin{document}

\title{{\Large \bf  On universal aspects of  the left-handed helix  region  \vskip 2.0cm }}

\vskip 1.0cm

\author{Martin Lundgren}
\email{Martin.Lundgren@physics.uu.se}
\affiliation{\large Department of Physics and Astronomy, Uppsala University,
P.O. Box 803, S-75108, Uppsala, Sweden}
\author{Antti J. Niemi}
\email{Antti.Niemi@physics.uu.se}
\affiliation{\large Department of Physics and Astronomy, Uppsala University,
P.O. Box 803, S-75108, Uppsala, Sweden}
\affiliation{\large
Laboratoire de Mathematiques et Physique Theorique
CNRS UMR 6083, F\'ed\'eration Denis Poisson, Universit\'e de Tours,
Parc de Grandmont, F37200, Tours, France}
\author{Fan Sha}
\email{fansha0559@gmail.com}
\affiliation{\large Department of Physics and Astronomy, Uppsala University,
P.O. Box 803, S-75108, Uppsala, Sweden}


\large
\begin{abstract}

\vskip 1.5cm
\large\baselineskip 0.8cm
\noindent
We  inspect the geometry
of proteins  by identifying their backbones  as framed 
polygons. We find that the left-handed helix region of the Ramachandran map for non-glycyl 
residues corresponds to an isolated and highly localized sector in the orientation
of the $C_\beta$ carbons, when viewed in a Frenet frame that is centered  at the corresponding
$C_\alpha$ carbons.
We show that this  localization in the orientation persists to $C_\gamma$ and $C_\delta$ carbons.
Furthermore, when we extend our analysis to the neighboring
residues we conclude that the  left-handed helix region  reflects
a very regular and apparently residue independent collective interplay of at least seven consecutive amino acids.
\end{abstract}



\maketitle
\vfill\eject
\baselineskip 0.8cm
\vskip 1.0cm
\begin{center}
{\bf I: INTRODUCTION}
\end{center}
\vskip 0.4cm
Asparagine (ASN) is the predominant non-glycyl residue in the left handed helix  ($\tt L$-$\alpha$) 
region of the Ramachandran map of folded proteins \cite{rama}, \cite{hov}.    According to  a prevailing  
view this reflects the presence of a localized but 
non-covalent attractive carbonyl-carbonyl interaction between the 
side-chain and  backbone  \cite{deane}, \cite{allen}, \cite{review}.  Such a carbonyl-carbonyl interaction can only be present
in ASN, aspartic acid (ASP), glutamine (GLN) and glutamic acid (GLU).  Indeed, the propensity of  ASP  that is structurally very
similar to ASN is clearly amplified in the $\tt L$-$\alpha$ region, while the somewhat lower propensity of GLN and GLU  
can be explained to be a consequence of steric suppression \cite{deane}. 
Consequently one may
suspect that  when located in the $\tt L$-$\alpha$ region, ASN and ASP residues should give rise  to atypical fold geometries.
 
ASN and ASP are also more frequently than any other amino acid subject to {\it in vivo} post-translational 
modifications including spontaneous nonenzymatic deamidation from ASN to ASP \cite{deami}
and racemization from  $\tt L$-ASP into $\tt D$-ASP \cite{race}. Since these processes are presumed to have consequences  
to  cellular and organismal  ageing \cite{deami}, \cite{deami2} and  they might also enhance the emergence of amyloid based neurodegenerative 
diseases \cite{prion}, \cite{deami2} there are several good reasons to search for those patterns in folded proteins  that appear to set these two residues apart 
from the rest.

In this article we apply  recently developed visualization  techniques  \cite{frenet} to analyze protein conformations
in the $\tt L$-$\alpha$ region. We are particularly interested in  the common  aspects of the ASN and ASP residues  
in this region.  But in lieu of the Ramachandran map which is topologically  a torus that has been projected onto the plane and as such is
subject to discontinuities, 
our approach  exploits  the visually more engaging two-sphere. 
For this we  interpret
a folded protein as a piecewise linear {\it framed } chain with vertices located at the $C_\alpha$ carbons \cite{frenet}. The 
framing can   be introduced in various different ways, and examples include the geometric Frenet frame \cite{hansonbook},
\cite{kuipers},  the geodesic Bishop frame \cite{bishop}, and
the  protein specific $C_\beta$ carbon frame that we obtain by utilizing the direction of the $C_\beta$ carbon along a protein backbone
to construct an orthonormal framing \cite{frenet}.   
This concept  of framed chains is  widely employed for example in aircraft and robot kinematics, stereo reconstruction and virtual reality 
\cite{hansonbook}, \cite{kuipers}.
In those applications different framings  correspond to different camera gaze positions, that one introduces  
for the purpose of extracting diverse and complementary information on various geometric
aspects and structural properties  of the system under investigation. However, thus  far this leeway  that can be enjoyed by deploying 
different frames has been sparsely applied in analyzing 
protein conformations.  Here we propose that the freedom in the choice of frames provides  a powerful and pristine tool for 
capturing universal aspects in the  geometry of folded proteins.

\section{Methods}

\noindent
{\bf A: Framing }
\vskip 0.4cm
The framing of a piecewise linear chain can be introduced
using the Denavit-Hartenberg \cite{dh} formalism that was  originally developed  in robotics but subsequently
applied extensively also in other disciplines. 
Here we resort to a variant that has been elaborated  in \cite{frenet}.  It  
utilizes the transfer matrix formalism  \cite{lattice} to
describe a protein with $N$ residues using the coordinates $\mathbf r_i$ of the backbone $C_\alpha$ 
carbons  ($i=1,...,N$); The coordinates can be downloaded from 
the Protein Data Bank (PDB) \cite{pdb}.
For each of the segments that connect the backbone $C_\alpha$ carbons  we compute  the unit 
length tangent vector $\mathbf t_i$, binormal vector $\mathbf b_i$ and normal vector $\mathbf n_i$ using
\[
{\bf t}_i = \frac{ {\bf r}_{i+1} - {\bf r}_i }{ | {\bf r}_{i+1} - {\bf r}_i |}
\ \ \ \ \& \ \ \ \ 
{\bf b}_i = \frac{ {\bf t}_{i-1} \times {\bf t}_i }{| {\bf t}_{i-1} \times {\bf t}_i|} 
\ \ \ \ \& \ \ \ \ 
{\bf n}_i = {\bf b}_i \times {\bf t}_i 
\]
The right-handed  triplet $(\mathbf n_i , \mathbf b_i , \mathbf t_i)$ constitutes the orthonormal 
Discrete Frenet frame (DF frame) for each residue along the backbone chain, with base  at the position of the vertex $\mathbf r_i$.  
At each vertex $i$,  a general orthonormal frame ($\mathbf e_1 , \mathbf e_2$) on the normal plane to $\mathbf t_i$ can be
obtained by rotating the DF frame around the tangent vector,
\begin{equation}
\left( 
\begin{matrix} 
{\bf e}_1  \\  {\bf e }_2 \\ {\bf t}
\end{matrix} \right)_{i} \ = \ 
\left(  \begin{matrix}  \cos \Delta & \sin \Delta & 0 \\
- \sin \Delta & \cos \Delta & 0 \\
0 & 0 & 1 \end{matrix} \right)_i 
\left( 
\begin{matrix} 
{\bf n}  \\  {\bf b } \\ {\bf t}
\end{matrix} \right)_{i} 
\label{GFF}
\end{equation}
The  parameters $\Delta_i$ specify the new frame at vertex $\mathbf r_i$. Since a change 
in $\Delta_i$ is a frame rotation around $\mathbf t_i$ 
it has no effect on the geometry of the curve. It only rotates the local frame $(\mathbf e_1, \mathbf e_2)$
at the $i^{th}$ vertex around $\mathbf t_i$. If we choose all $\Delta_i \equiv 0$ we get the DF frame at each vertex, while
for non-vanishing choices of $\Delta_i$  we obtain alternative frames. 
In \cite{frenet} it has been shown that the generic set of frames is subject  to the following  generalized DF equation
\begin{equation}
\left( 
\begin{matrix} 
{\bf e}_1  \\  {\bf e }_2 \\ {\bf t}
\end{matrix} \right)_{i+1}
= \  {\mathcal R}_{i+1,i} 
\left( \begin{matrix} {\bf e}_{1} \\  {\bf e }_{2} \\ {\bf t} \end{matrix} \right)_i
\label{DFE}
\end{equation}
Here the transfer matrix that determines the frame at site $i+1$ in terms of the frame at site $i$ is
\[
{\mathcal R}_{i+1,i}   = \exp\{ \kappa_{i+1,i} ( T^2 \cos \Delta_{i+1} - T^1 \sin \Delta_{i+1} ) \}
\]
\[
\cdot \exp \{
( \tau_{i+1,i} + \Delta_{i} - \Delta_{i+1} ) T^3 \}
\]
The $T^a$ $(a=1,2,3)$  are the adjoint SO(3) Lie algebra generators, and $\kappa_{i+1,i}$ and $\tau_{i+1,i}$ are the geometrically 
determined bond and torsion angles, defined as shown in Figure 1.
\begin{figure}[!hbtp]
  \begin{center}
    \resizebox{6cm}{!}{\includegraphics[]{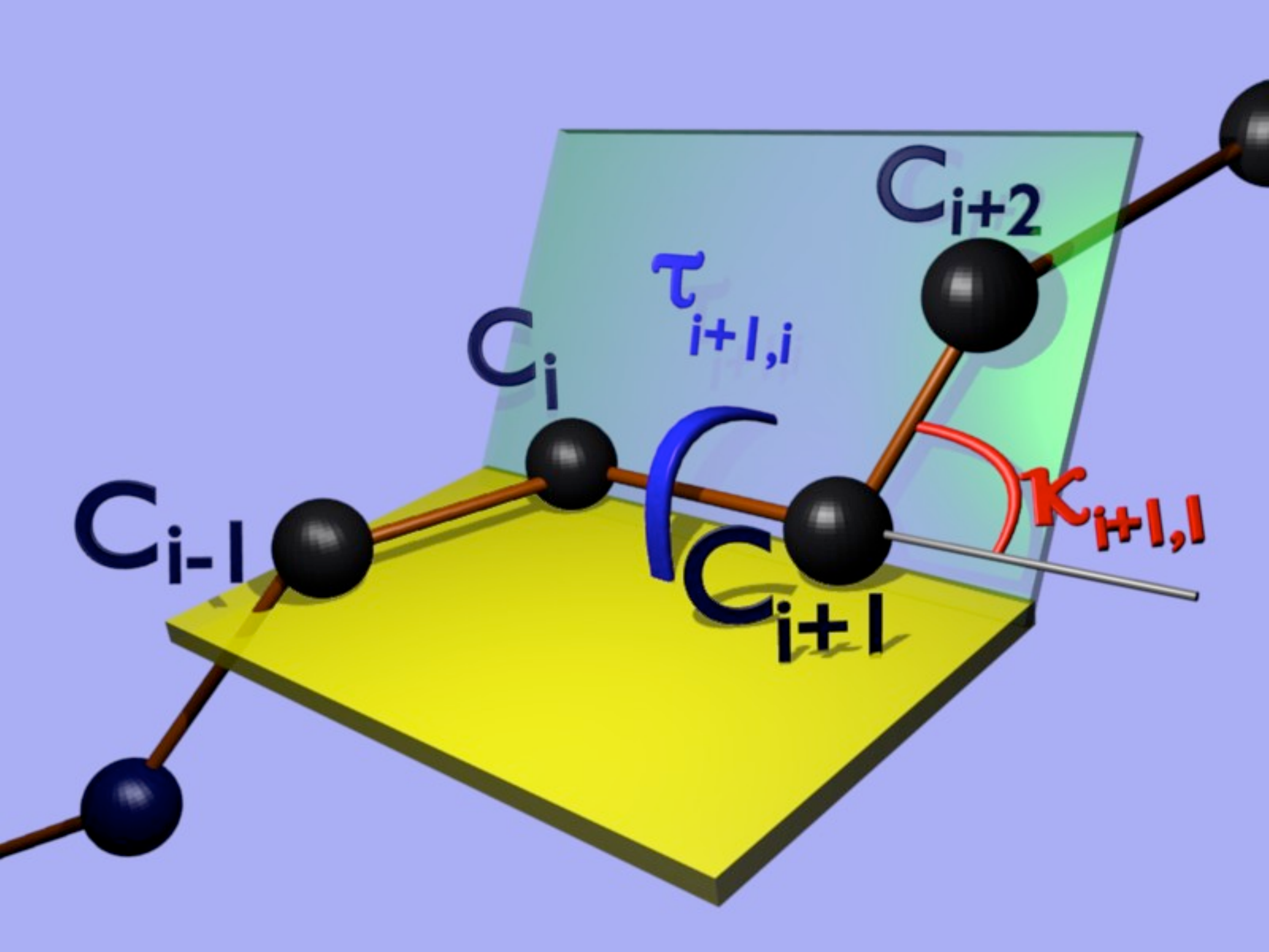}}
    \caption{Definition of bond $\kappa$ and torsion $\tau$ angles in terms of the backbone $C_\alpha$ carbons. }
    \label{fig:figure-1}
  \end{center}
\end{figure}
Note that according to (\ref{DFE}), the bond and torsion angles are link variables. In particular, the definition of the bond angle 
involves three vertices while the definition of the torsion angle involves a total of four vertices. 

We have employed the generalized DF equation (\ref{DFE}) to inspect  the structure of folded  proteins.  As a data set 
we have utilized all those proteins that are presently in PDB,  with an overall resolution better
than 2.0 \.Angstr\"om. There is no additional curation or data pruning.  As a control set  we have used
the highly curated version v3.3 Library of chopped PDB files for representative  CATH domains \cite{cath}. Since the conclusions we 
draw from these two data sets are parallel, we only describe here  the results for the  first one in detail. 

\vskip 0.4cm

\noindent
{\bf B:  Backbone Visualization}
\vskip 0.4cm

We start by describing the visualization of the  (frame independent) directions of the  tangent vectors $\mathbf t_i$ along the 
backbone.  For this we note  that with the base of $\mathbf t_i$ at the location 
of the corresponding $C_\alpha$ carbon, its tip 
determines a point on the surface  of a unit two-sphere. The location of this point is described
by the bond angle as the latitude, and the torsion angle as the longitude of the two-sphere.
Since the $\mathbf t_i$ are frame independent,  the traces of their tips on the two-sphere provides 
frame independent information of the backbone geometry.  For visualization of the two-sphere, we can
stereographically project it
 from south-pole $(\kappa_i \equiv \kappa_{i, i+1} = \pi)$ onto the two dimensional plane. This leads us to  the angular 
 distributions that we have displayed in Figure 2,  and in each  of the four classes 
 in this Figure we have used  the prevalent  PDB identification of  the ensuing structures.

\begin{figure}[!hbtp]
  \begin{center}
    \resizebox{8.0cm}{!}{\includegraphics[]{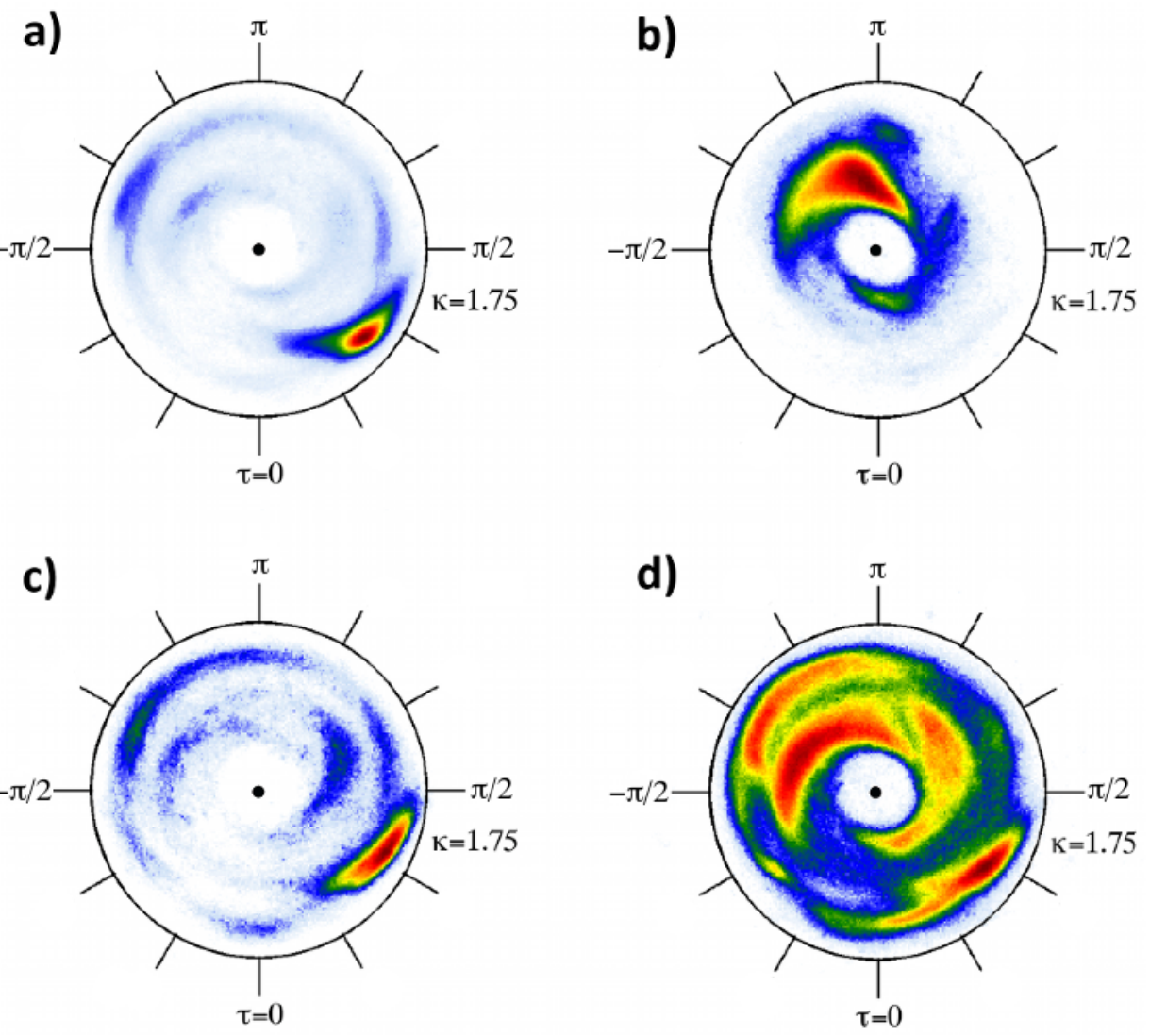}}
    \caption{The four major protein structures: a) 
$\alpha$-helices,  b) $\beta$-strands,   c) $3/10$-helices and d) loops  according to
PDB classification in our data set.
In each Figure the center of the annulus is  the north-pole of the two-sphere  where the bond (latitude) angle
$\kappa = 0$ so that the two consecutive unit tangent vectors $\mathbf t_i$ and $\mathbf t_{i+1}$ are parallel. 
The bond angle then measures distance from the center of the annulus so that 
the south pole  where $\kappa=\pi$ corresponds to points at infinity on the plane, where  $\mathbf t_i$ and $\mathbf t_{i+1}$ become
anti-parallel.  The torsion {\it i.e.}  longitude 
angle $\tau\in [-\pi,\pi]$  increases by $2\pi$ when we go around the center of the annulus in counter-clockwise direction. 
The color coding in all our Figures  increases from white to blue to green to yellow to red  and describes the relative number of conformations in 
PDB in a $\log$-squared scale. } 
    \label{fig:figure-2}
  \end{center}
\end{figure}

The four maps in Figure 2 portray all the essential features  of the Ramachandran map. 
But there are also important differences that makes it profitable to utilize these maps both in lieu of and in combination with the Ramachandran map.
The predominant feature in Figures 2 is  that the 
PDB data is concentrated in an annulus which is  located roughly in the range $ \kappa \approx (1, \pi/2)$. The exterior of the 
annulus (roughly  $\kappa > \pi/2$) is an excluded region that describes  conformations with steric clashes. The  
interior (roughly $\kappa < 1$) 
is  sterically allowed but excluded as 
long as proteins remain in the collapsed phase; The interior region becomes occupied when we cross  the  $\Theta$-point 
and proteins  assume their unfolded conformations.  The  loops also appear to have a slightly higher tendency to bend towards left  {\it i.e.}  $\tau < 0$.
Note that in the Figures for  $\alpha$, $\beta$ and $3/10$  the blue regions correspond to residues whose
present PDB classification are not  consistent with our  computed $(\kappa,\tau)$ values; Such issues have also been raised in \cite{hov}. 

Moreover, the Figure 2 reveals that the PDB data displays innuendos of various 
underlying reflection symmetries: In the  Figure 2d (loops) there is a clearly visible 
mirror of the standard right-handed $\alpha$-helix region, located in the vicinity of the outer 
rim with $\kappa \approx 1.5  $ and with torsion angle close to the value $\tau \approx -2\pi/3$. A helix in this regime would be 
left-handed and tighter than the standard $\alpha$-helix. There is also
a clear mirror structure in the Figure 2b for $\beta$ strands,  the standard region is $(\kappa, \tau) \approx ( 1,\pi)$  and its less populated 
mirror is located around
$(\kappa, \tau) \approx (1,0)$. The mirror symmetry between the ensuing extended regions persists in the Figure 2d for loops. 

Finally, in the Figure 2d
we observe  a small elevated (yellow)
region in the vicinity of $(\kappa, \tau) \approx (1.5, -\pi/3)$. 
This is the region of helices that are spatial left-handed mirror images of the standard $\alpha$-helices. There is also 
a (slightly) elevated (green) mirror of this region around $(\kappa, \tau) \approx (1.5, 2\pi/3)$.  This is like 
the  $(\kappa, \tau)  \approx (1.5\, ,  -2\pi/3)$ mirror of the 
standard right-handed $\alpha$ helices.

\vskip 0.6cm

\noindent
{\bf C:  Side-Chain Visualization}
\vskip 0.4cm

In Figure 3 (top) we display the orientations of the $C_\beta$ carbons in the Frenet frames of the $C_\alpha$ carbons 
{\it i.e.} all $\Delta_i = 0$ in (\ref{GFF});
Recall that
a $C_\beta$ carbon is present in all non-glycyl residues. In this Figure the $C_\alpha$ carbons
are at the center of the sphere.   Consequently the Figure 3 (top) shows directly  the locations of the $C_\beta$ carbons as seen by an observer
who traverses the $C_\alpha$ backbone with a gaze orientation that is determined by  the  DF  frame.   
We find it  remarkable that in this frame the directions
of the $C_\beta$ carbons are subject to  only very small  nutations. The additional  feature is the presence of the  highly localized, isolated  island.
 \begin{figure}[!hbtp]
  \begin{center}
    \resizebox{5.cm}{!}{\includegraphics[]{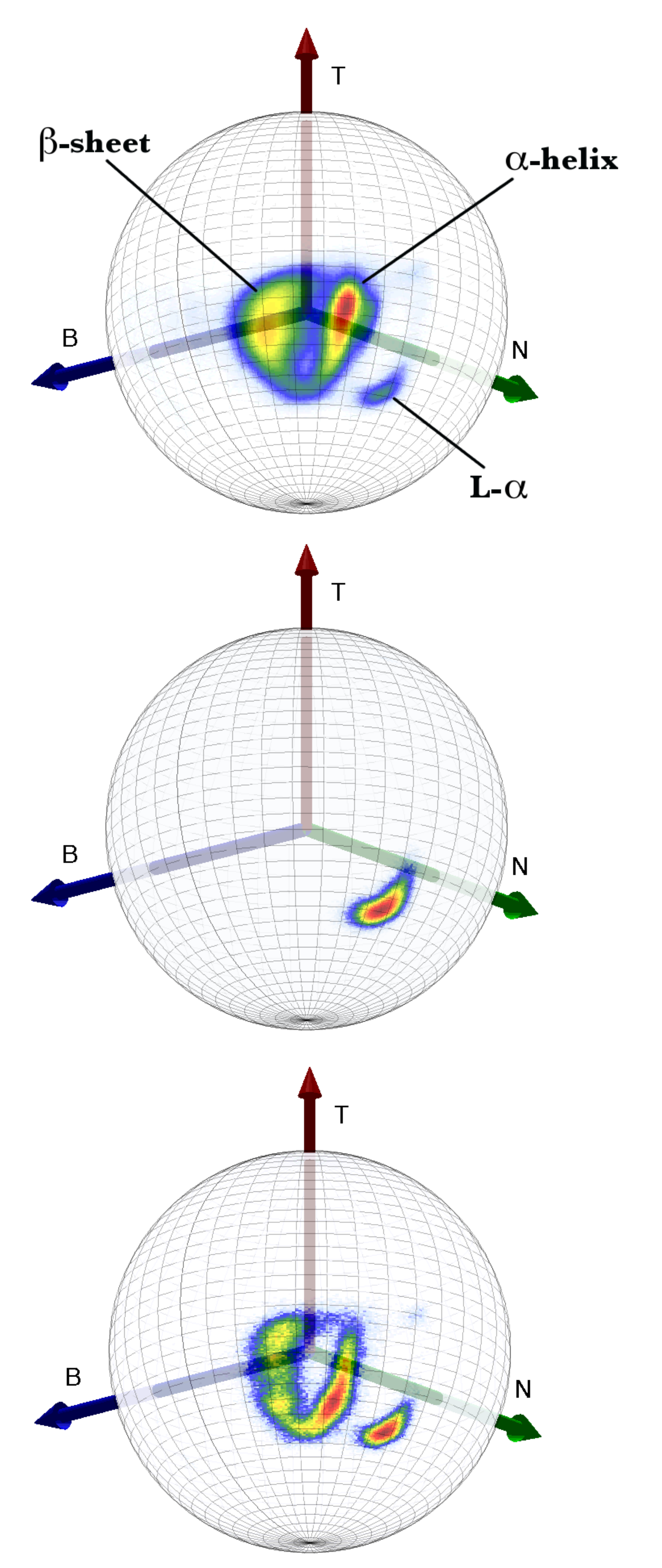}}
    \caption{The DF frame directions of the $C_\beta$ carbons.
    On top all residues in our data set including ASN. In the middle we display only the $\tt L$-$\alpha$
    region of the Ramachandran map. On bottom we display  only those ASN  that are in a loop in PDB classification.  The color coding is
    proportional to the ensuing propensity.  }
    \label{fig:figure-3}
  \end{center}
\end{figure}
Indeed, we have found that  for non-glycyl residues  this  isolated island coincides  with the  $\tt L$-$\alpha$ region of the Ramachandran map. 
This is shown in Figure 3 (middle) where we display the direction of the $C_\beta$ carbons solely  for those residues that are in the $\tt L$-$\alpha$  Ramachandran region.
Finally, in the Figure 3 (bottom) we display the DF frame distribution of the $C_\beta$ carbons for those ASN that are located
in loops only, according to PDB classification.  The relatively
high propensity of ASN  in the $\tt L$-$\alpha$ island  is prominent. 
In the sequel we shall concentrate our attention solely on the isolated  
$\tt L$-$\alpha$ island in Figure 3 (middle).

Figure 4  describes the propensity of different amino acids  in the $\tt L$-$\alpha$ island of Figure 3.   
This Figure confirms the high propensity of ASN (N) that is already visible in Figure 3 (bottom).  We find that ASP (D) has also 
relatively high propensity. But  the propensity of  histidine (H) is practically  equal in our data. Furthermore, several non-carbonylic
amino acids have a higher propensity than GLU (E).   
The $\beta$-branched isoleucine (I), valine (V) and threorine (T)  all have
clearly suppressed propensities and proline (P) is practically absent, presumably reflecting the presence of steric constraints \cite{deane}, \cite{hov}, 
\cite{review}.

 \begin{figure}[!hbtp]
  \begin{center}
    \resizebox{8.5cm}{!}{\includegraphics[]{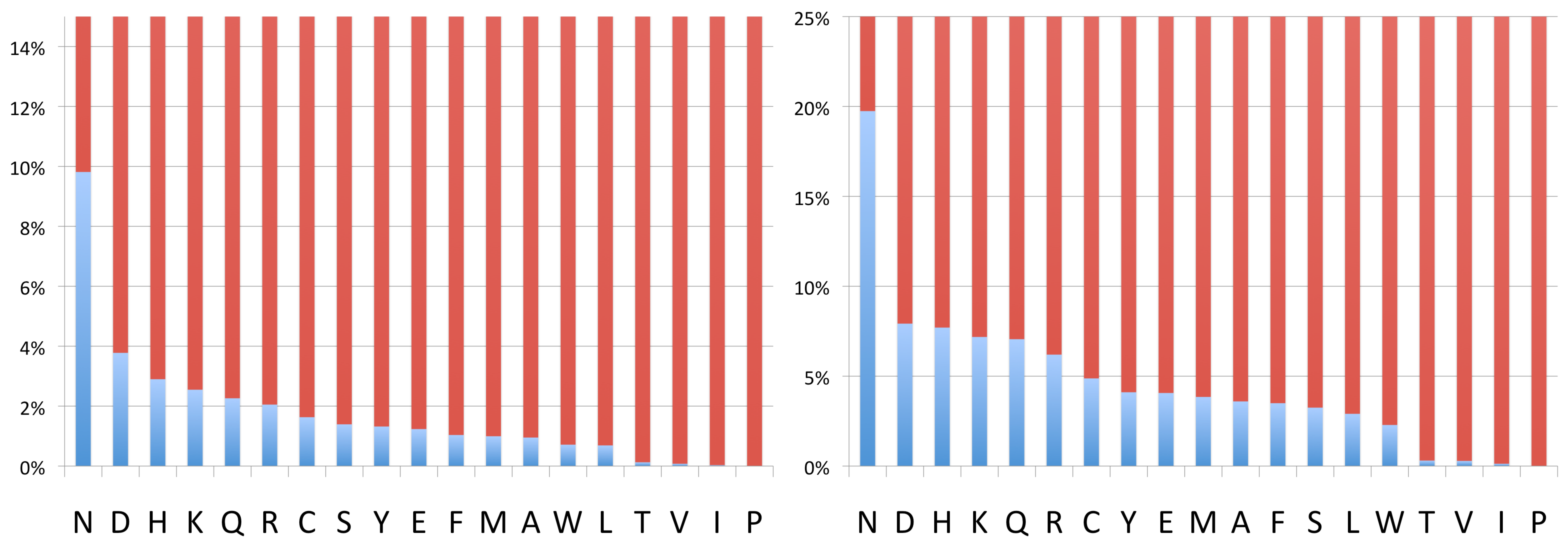}}
    \caption{The propensity of non-glycyl  residues in the $L$-$\alpha$ island of Figure 3. In the left we display the result for
    all amino acids in our entire data set, and in the right for those in our data set that are classified as loops in PDB. The propensity of
     carbonylic ASN (N) is clearly enhanced in both cases.  But in both cases the similarly carbonylic ASP (D) has about the same propensity 
     with the non-carbonylic HIS (H), and the  carbonylic GLU (E) is relatively quite suppressed. 
       }
    \label{fig:figure-4}
  \end{center}
\end{figure}

We have also analyzed the directions of the $C_\gamma$ carbons,  as seen by a $C_\alpha$  observer
in the DF frame.  The  result presented in Figure 5  reveals 
that at the level of $C_\gamma$  the  data points in the single $\tt L$-$\alpha$ island of Figure 3 remain highly localized even though it now
becomes divided into two 
separate islands. There is a putative  {\it gauche}+  ($g$+) island
where around 70$\%$ of the residues  in the $L$-$\alpha$ island are located, and a putative  {\it trans} 
island for the rest; We do not see any  putative  {\it gauche}- region. The amino acid  propensities of these two islands is displayed in Figure 6.
ASN is the most populous  in both $C_\gamma$ islands. However, the propensity of ASP is elevated
only in  {\it trans} island. In the  {\it g}+ island both non-carbonylic HIS  (H) and LYS (K) and even the carbonylic GLN (Q) have a 
higher propensity than ASP.

\begin{figure}[!hbtp]
  \begin{center}
    \resizebox{5cm}{!}{\includegraphics[]{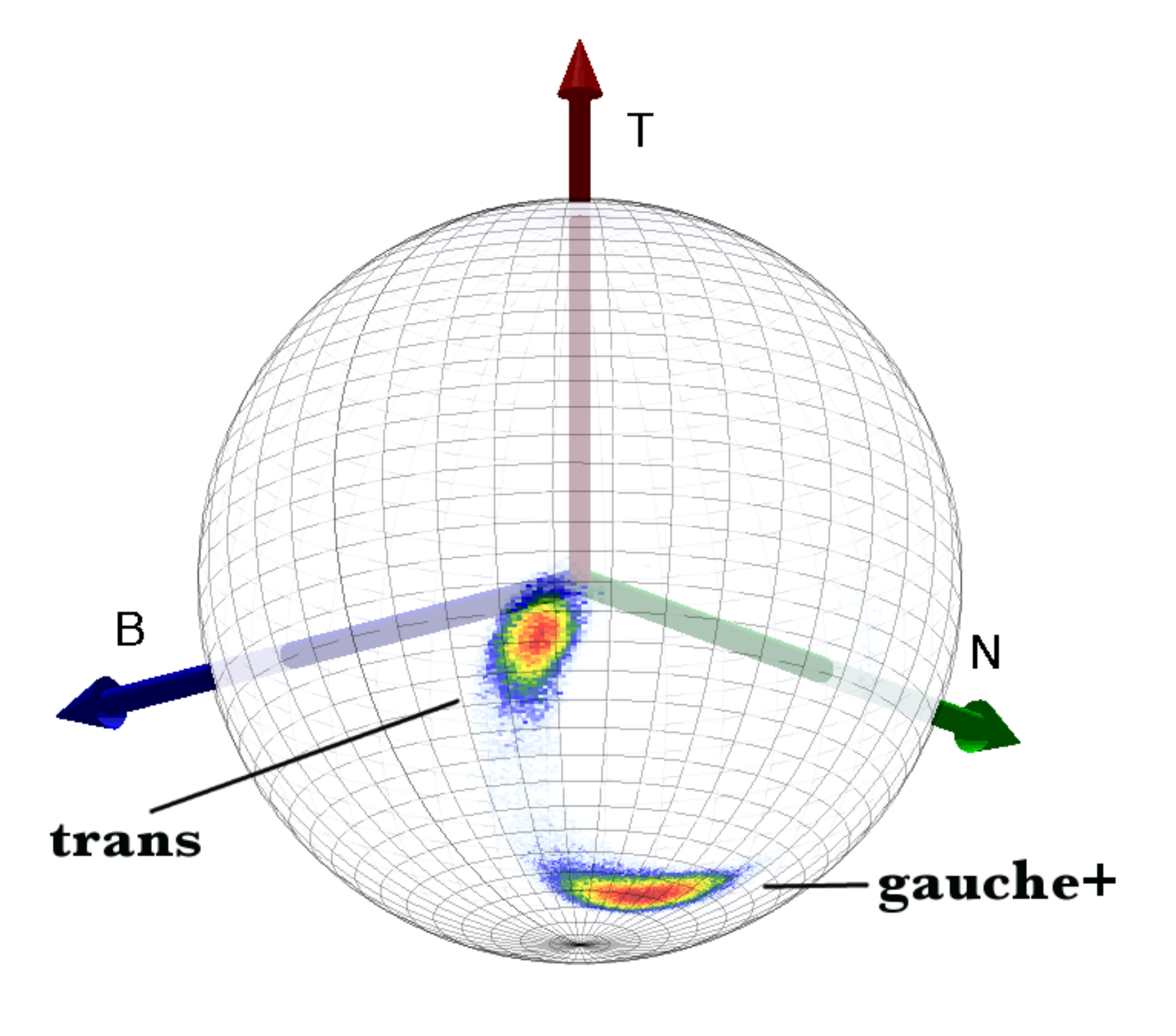}}
    \caption{The DF frame directions of $C_\gamma$  carbons for residues  located in the  $\tt L$-$\alpha$ island.
       }
    \label{fig:figure-5}
  \end{center}
\end{figure}

\begin{figure}[!hbtp]
  \begin{center}
    \resizebox{8cm}{!}{\includegraphics[]{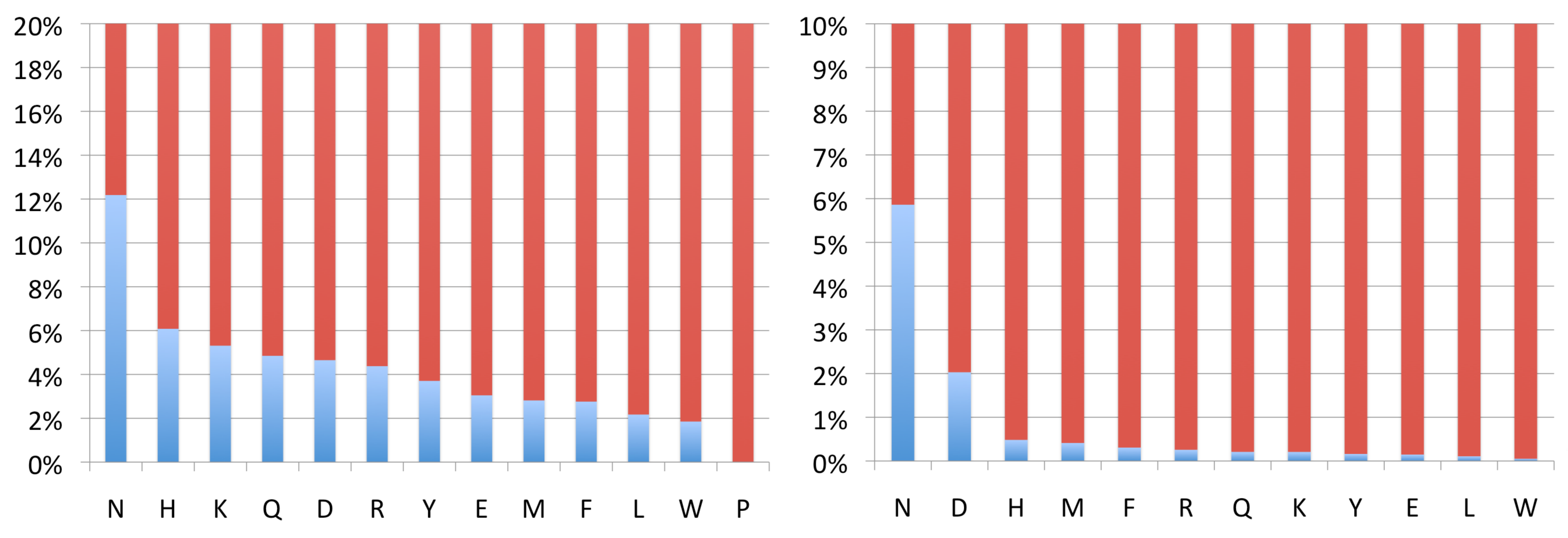}}
    \caption{The propensity of different amino acids in the putative {\it g}+ island (left) and {\it trans} island (right)}
    \label{fig:figure-6}
  \end{center}
\end{figure}
\noindent

 In Figure 7 we plot the percentage of different amino acids as they appear in our data set in the two 
$C_\gamma$ islands. Note that around $43\%$ of residues  in the putative $\it g+$ island  are 
non-carbonylic, while in the putative $\it trans$ island the number is close to $12\%$. 

\begin{figure}[!hbtp]
  \begin{center}
    \resizebox{8.5cm}{!}{\includegraphics[]{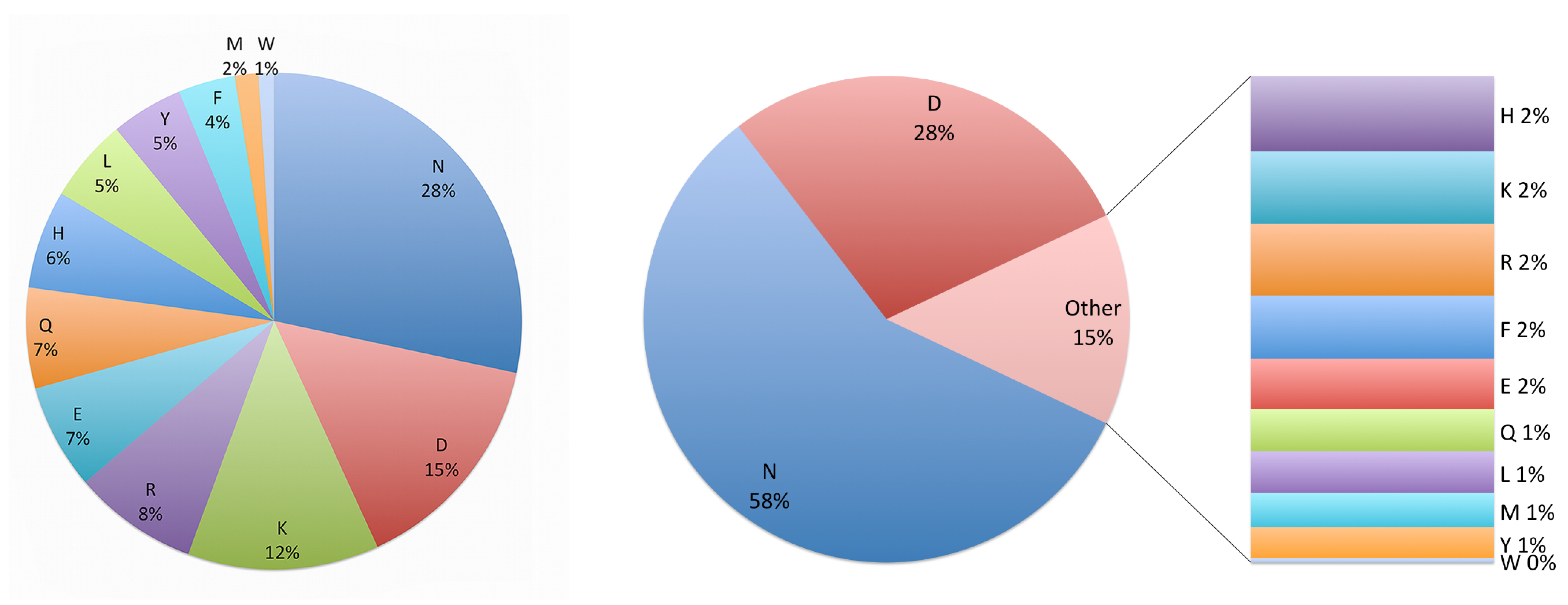}}
    \caption{The relative number of different amino acids in the putative $\it g+$ (left) and $\it trans$ (right) $C_\gamma$-islands.}
    \label{fig:figure-7}
  \end{center}
\end{figure}

In Figure 8 we  plot the $C_\delta$ carbons in the  DF frame of the $C_\alpha$ carbons.  Since  ASN (and ASP)  has no
$C_\delta$ carbon, we display instead the direction of the
side-chain $O$ atom for ASN, and the result is shown in Figure 9. 
\begin{figure}[!hbtp]
  \begin{center}
    \resizebox{7.5cm}{!}{\includegraphics[]{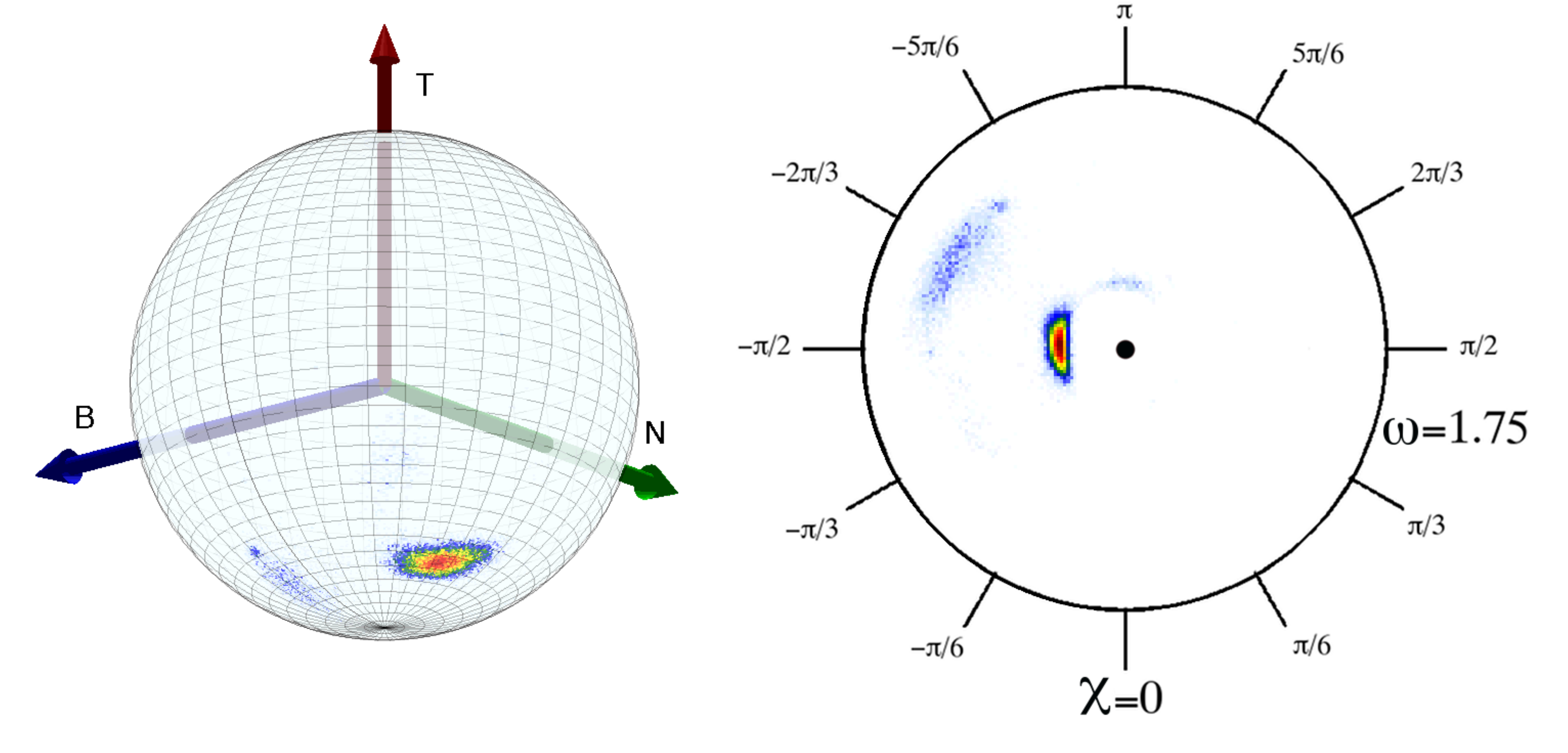}}
    \caption{The directions of the  $C_\delta$-carbons in the DF frame of the  $C_\alpha$  carbons on the two-sphere (left), 
    and on  stereographically projected $C_\beta$ frame (right).}
    \label{fig:figure-8}
  \end{center}
\end{figure}

From Figure 8 we observe that the directions of the $C_\delta$ continue to be highly localized, independently of the type of
amino acid. We also observe the formation of a second but relatively  
weakly occupied island at larger values of the latitude angle and with longitude
angle $\chi \sim -2\pi/3$, clearly visible  in Figure 8 (right).
At the moment we do not have a basis to conclude whether this is a real effect or simply a reflection of problems in the experimental data.
\begin{figure}[!hbtp]
  \begin{center}
    \resizebox{4.2cm}{!}{\includegraphics[]{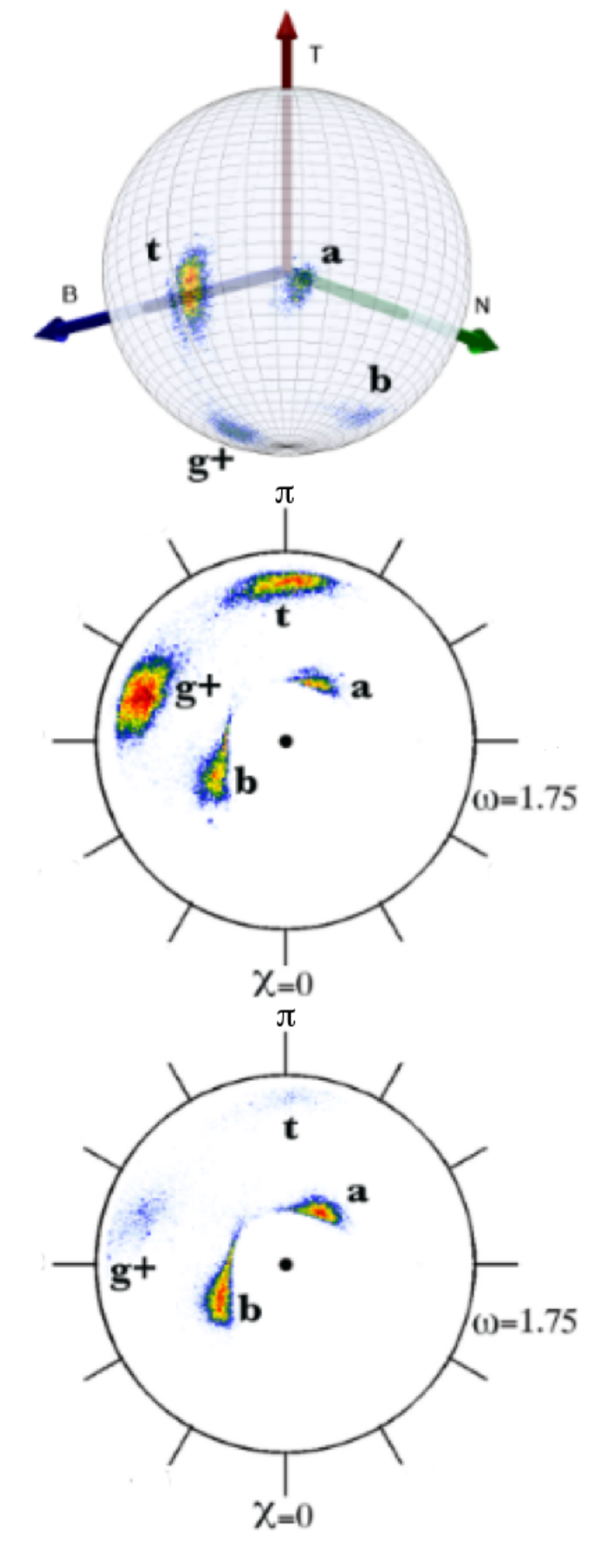}}
    \caption{On top, the directions of the  side-chain $O$ atoms of ASN  in the DF frame of the  $C_\alpha$  carbons.  In the middle, the same as on top
    but on stereographic projection and in the  $C_\beta$ frame. Bottom, 
    the directions of the  side-chain $N$ atoms of ASN  in the same $C_\beta$ frame as in middle suggesting that the correct identification 
    of {\bf a} and {\bf b} regions in the top and middle figure is $N$.  The identifications follow PDB. ({\bf t} is {\it trans} and {\bf g+} is {\it gauche+})}
    \label{fig:figure-9}
  \end{center}
\end{figure}

In Figure 9 we display the $O$ atoms of the ASN side-chain  according to PDB identification 
in the DF frame of the $C_\alpha$-carbons. We note that the two $C_\gamma$ islands appear to become divided into four distinct but still highly 
localized islands. However, it is well known that the identification between the ASN side-chain $O$ and $N$ can be very difficult \cite{errors}, 
and thus we have displayed in Figure 9 (bottom) the $N$ atoms  according to PDB identification
as well. By comparing the Figures 9 (middle) and (bottom) we conclude that most likely
the two inner-most islands denoted {\bf a} and {\bf b}  in Figure 9 describe $N$ instead of $O$ atoms.

\section{Results and Discussion}

Previously it has been proposed that in the case of ASN and ASP, the $\tt L$-$\alpha$ Ramachandran region 
is stabilized by a non-covalent attractive interaction between the side-chain and backbone carbonyls  \cite{deane}, \cite{review}.  Here we find
that in the  DF frame and  {\it  independently} of the amino acid,  the  ensuing side-chain  $C_\beta$ carbons all point  
 to the same direction that we have denoted $\tt L$-$\alpha$ in Figure 3. We have shown
that this region does also coincide with the  non-glycyl $\tt L$-$\alpha$ region of the Ramachandran map.
 Furthermore,  
the strong localization in the direction  continues to persist  when we  extend our analysis to the $C_\gamma$ and $C_\delta$ carbons of the
$\tt L$-$\alpha$  residues, the results 
are displayed in Figures 6 and 9 respectively;  In the case of ASN (ASP) where there is no $C_\delta$ carbon we utilize
 the side chain $O$ and $N$ atoms instead and as shown in Figure 10 the results are very similar.  
This strong localization in the directions and in particular  its apparent  residue independence suggests that 
the presence of the $\tt L$-$\alpha$ island  is  associated with some relatively residue independent
structural property  of the protein backbone. 
 
According to \cite{deane},  in the case of ASN and ASP the backbone oxygen atom has a special r\^ole.  If so, it  should
somehow be reflected  in the ensuing backbone conformations.
In order to scrutinize this,
we have inspected the orientations of all  backbone $O$ atoms in our data set, in a group of residues where the $i^{th}$ one  is
located in the island.   The result is shown in Figure 11:  We find that there is a very strong localization in the directions of the backbone $O$ atoms,
and this localization is residue {\it independent }  and extends itself over at least four different residues: The $O$ atoms in both the $\tt L$-$\alpha$ site 
$i$ and at the preceding site $i-1$ are highly localized into a single direction, while  for both the site $i-2$ and the site $i+1$ we identify
three closely located directions, that are presumably related to the three available  {\it trans/gauche} conformations.

Indeed, it appears that only very few backbone geometries are 
accessible in the vicinity of a residue that is located in the $\tt L$-$\alpha$ island.  Furthermore, since  the regime that extends from the 
$(i-2)$nd to  the $(i+1)$st site  involves 
three sets of curvature and torsion angles, each of them defined in terms of three {\it resp.} four residues we conclude that the backbone 
geometries reflect the collective interplay of at least up to seven different sites along the backbone. 
\begin{figure}[!hbtp]
  \begin{center}
    \resizebox{4.5cm}{!}{\includegraphics[]{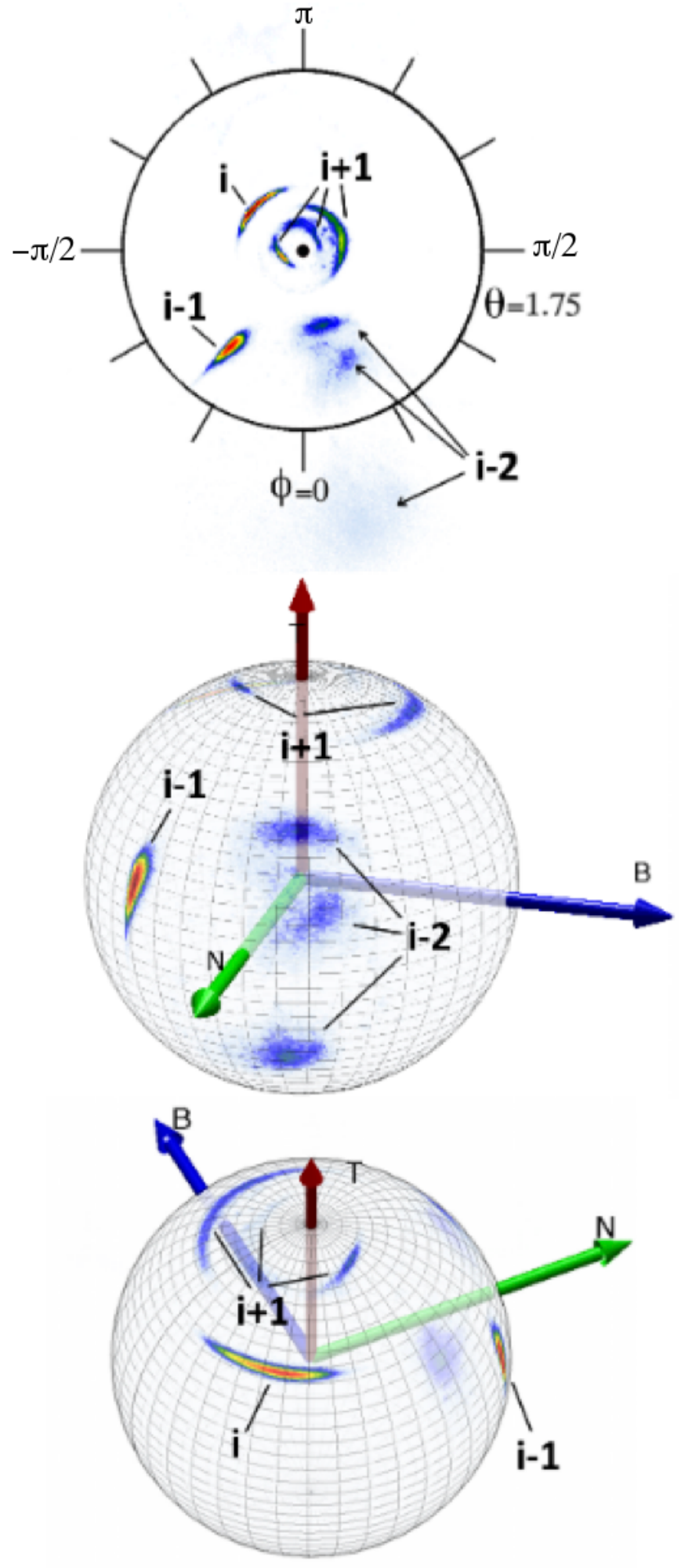}}
    \caption{ The orientations of backbone $O$ atoms around the site $i$ that is located in the  $\tt L$-$\alpha$ island, in
    the $C_\alpha$ discrete Frenet frame of the $i^{th}$ central $C_\alpha$-carbon.  For the  $i^{th}$ and $(i-1)$st atom only one position
    appears to be available while the $(i+1)$st and $(1-2)$nd atoms each have three available  ({\it trans/gauche}) positions.  
    The angle $\phi$ is measured from the $\bf N$
    axis.  }
    \label{fig:figure-10}
  \end{center}
\end{figure}

To further inspect the universality of the $\tt L$-$\alpha$ island,  we consider the distribution of the backbone bond and torsion angles  
that are attached to a $C_\alpha$ carbon when
a residue is located in the $\tt L$-$\alpha$ island. The result is shown  in Figure 11 separately for ASN and ASP, and for the remaining
 non-glycyl  amino acids.
\begin{figure}[!hbtp]
  \begin{center}
    \resizebox{8.cm}{!}{\includegraphics[]{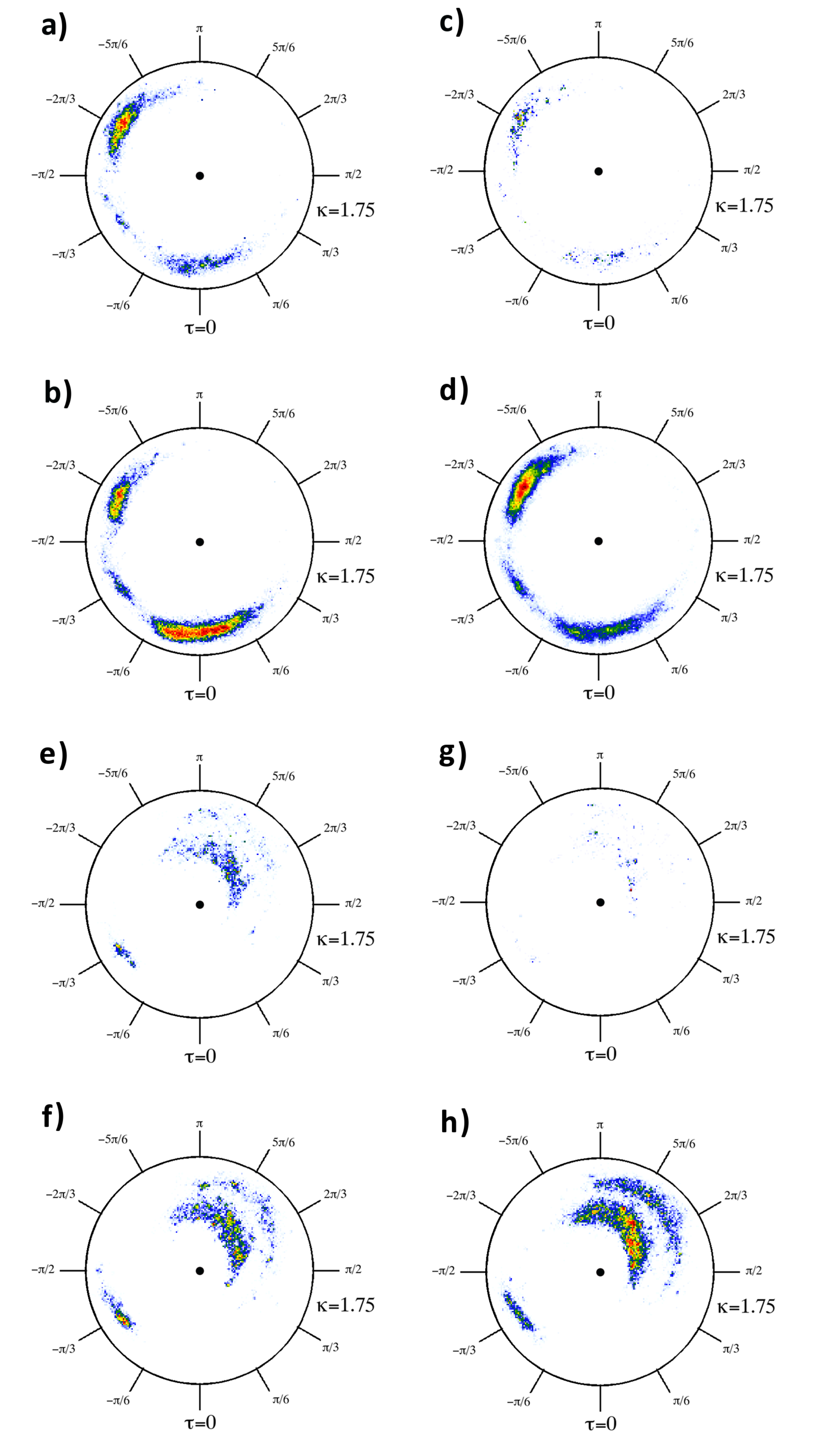}}
    \caption{The $(\kappa,\tau)$ distributions for backbone links that are attached to a $C_\alpha$ carbon with residue in the $\tt L$-$\alpha$ island. 
    Separately for 
    ASN and ASP, and for the rest. On left column  the $C_\alpha$ carbons 
    in the case where the corresponding $C_\gamma$ carbon is in the  {\it trans} island,
    on right for those where the $C_\gamma$ is in the $g$+  island. First row is for link that precedes 
    either ASN or ASP. Second row is
    for link preceding  any other  non-glycyl  amino acid.    Third row is for link following either ASN or ASP.  Fourth row is for 
    link following the others. }
    \label{fig:figure-11}
  \end{center}
\end{figure}
We find no essential difference between  the different residues, nor do we find any essential difference between the {\it trans} and $g$+ islands. 
Moreover, we observe the following general pattern: For  the backbone $C_\alpha$ link that
precedes the $\tt L$-$\alpha$ island,  three different  regions on the $(\kappa, \tau)$ plane are probable. These are the regions
that we have denoted with {\bf a}, {\bf b} and {\bf c} respectively  in  the Figure 12-left; In this Figure we have combined all the  
data that are displayed separately in the parts {\bf a, b, c} and {\bf d} of Figure
11.  After the  $\tt L$-$\alpha$ island  there are also three different regions that are probable. We denote  these regions  
with letters {\bf b} and {\bf d} and {\bf e} respectively in the Figure 12-right,  now combining  the data in parts {\bf e, f, g, h} in Figure 11.  
Note that the regions {\bf b} in the two parts of Figure 12 practically coincide. 
\begin{figure}[!hbtp]
  \begin{center}
    \resizebox{8.5cm}{!}{\includegraphics[]{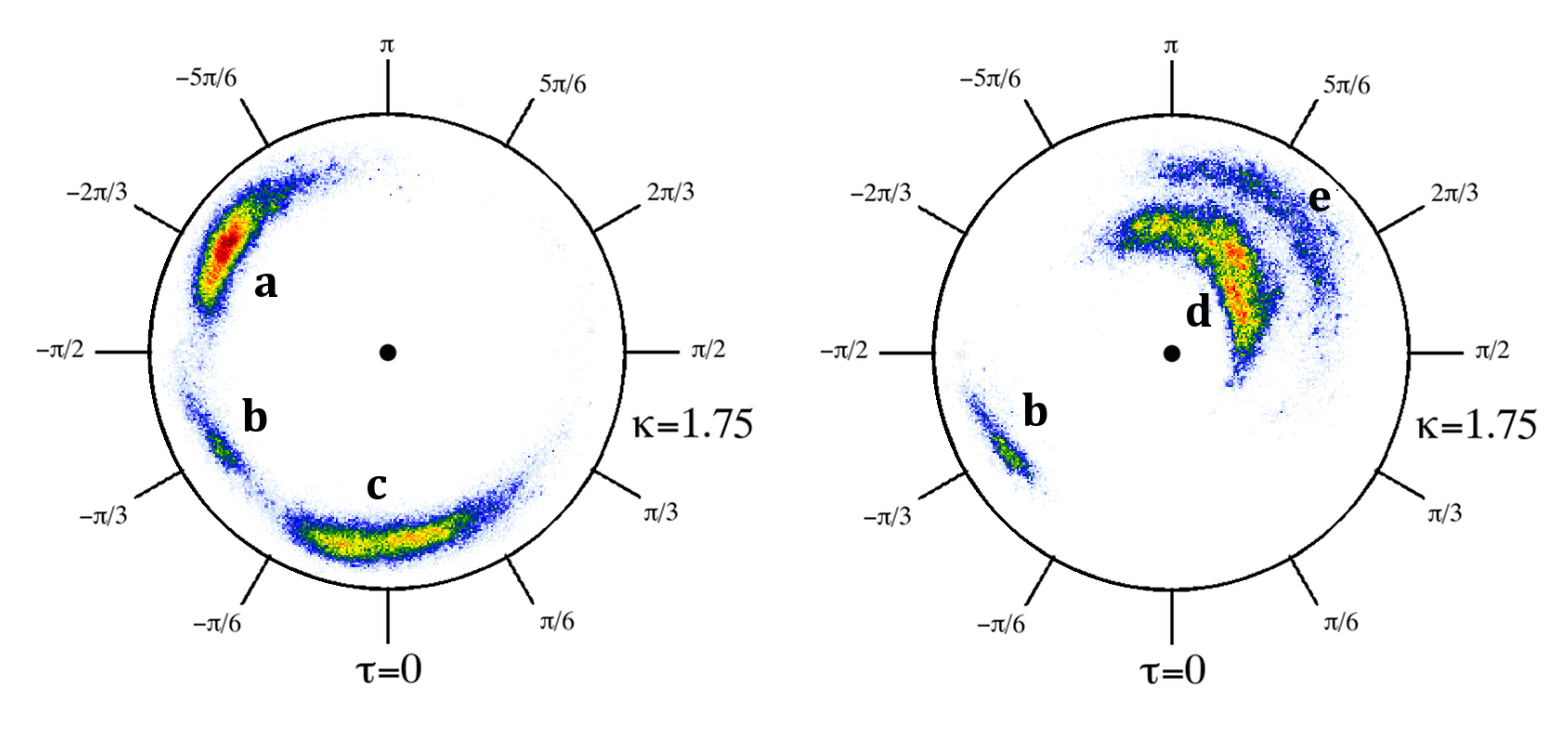}}
    \caption{The $(\kappa,\tau)$ distributions for all backbone links that are attached to a residue in the $\tt L$-$\alpha$ island. On the left for the link preceding
    the residue in the $\tt L$-$\alpha$ island, on the right following the residue in the $\tt L$-$\alpha$ island.}
    \label{fig:figure-12}
  \end{center}
\end{figure}
By inspecting the protein structures in our data set we conclude 
that the presence of a residue in the $\tt L$-$\alpha$ island causes the following phenomenologically verified {\it selection rules} in Figure 12:  

\vskip 0.4cm

$\bullet$ The region {\bf a}   can only precede regions {\bf d} and {\bf e}.  

$\bullet$ Both regions {\bf b} and {\bf c} can be followed by any of the three regions

\hskip 0.25cm {\bf b}, {\bf d} and {\bf e}.   

\vskip 0.4cm
\noindent
Furthermore, we find that 
\vskip 0.4cm

$\bullet$ The  residue preceding either {\bf a} or {\bf c}  is not
located in the $\tt L$-$\alpha$ 

\hskip 0.25cm island. 

$\bullet$ Both the residue preceding  and following {\bf b} can be  located in 

\hskip 0.25cm the  $\tt L$-$\alpha$ island. 

$\bullet$ If the two residues following {\bf c}  are both  in the  $\tt L$-$\alpha$ island, the

 \hskip 0.25cm first  residue  connects {\bf c} to {\bf b} and the second connects  {\bf b} to {\bf b}. 
 
\vskip 0.4cm

We remind that since the  region {\bf b} has the same curvature angle as the standard $\alpha$-helix region and
since the torsion angles are  equal in magnitude but have an opposite  sign, a repeated structure in {\bf b} is the
right-handed mirror image of the standard $\alpha$-helix, {\it this} is truly the region of  {\it left-handed $\alpha$-helices}.

We have also found that there appears to be 
four main trajectories that are followed by the orientations of those $C_\alpha$ carbons that surround  a residue located
in the  $\tt L$-$\alpha$ island. The result is shown in  the schematic Figure 13, where the pink arrows correspond to residues in the island;
Recall that the curvature and torsion angles are link variables, they connect two  $C_\alpha$ carbons according to (\ref{DFE}).
\begin{figure}[!hbtp]
  \begin{center}
    \resizebox{8.cm}{!}{\includegraphics[]{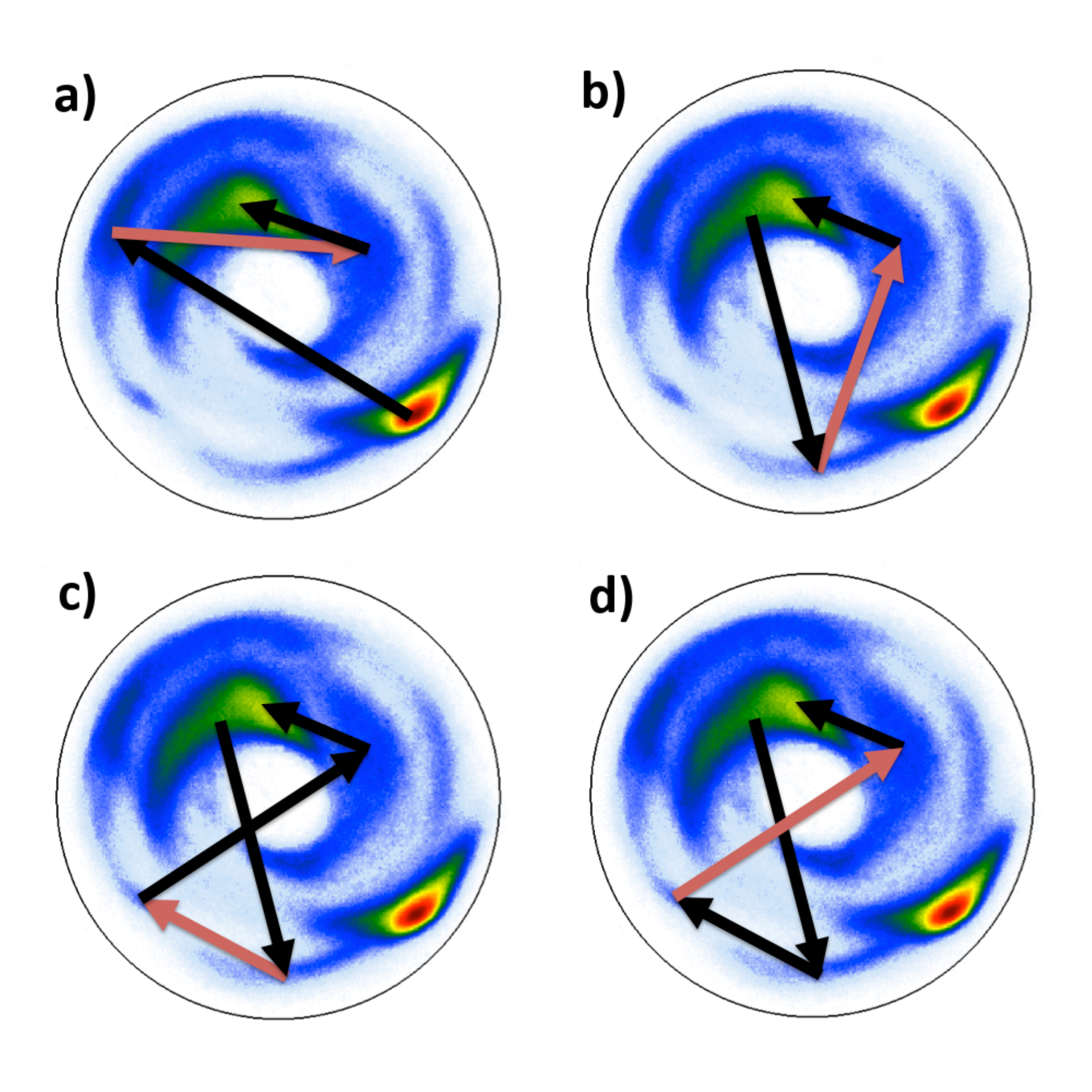}}
    \caption{ Four different trajectories through a residue in the $\tt L$-$\alpha$ island that are common in our data set.  In each case the pink lines denotes 
    the transition caused by  the residue in the $\tt L$-$\alpha$  island. The trajectory  {\bf a} described a turn from $\alpha$-helix region to $\beta$-strand region,      
    and the remaining ones start and end in the   $\beta$ region; these include $\beta$-turns.  }
    \label{fig:figure-13}
  \end{center}
\end{figure}

In Figure 13a, the first residue takes us  away from an $\alpha$-helix region  to the region
{\bf a} in the Figure 12 left (black arrow).  This is followed by a residue in the $\tt L$-$\alpha$ island, that takes us to the region {\bf d} in
the Figure  12 right (pink arrow). Finally, there is  a transition to the $\beta$-strand region (black arrow). 

The second trajectory in Figure 13b starts from the $\beta$-strand region with a residue that  takes it into region {\bf c} in Figure 12 left. 
The following  residue  that is located in the $\tt L$-$\alpha$ island then causes a transition into region {\bf d} in Figure 12 right (pink arrow).
This is followed by a transition back to a $\beta$-strand region. 

The third trajectory that we have described in Figure 13c starts from the $\beta$-strand region and proceeds to region {\bf  c} in Figure
12 left.  From there  the trajectory proceeds to region {\bf b}
in Figure 12 left,  with the transition caused by a residue in the $\tt L$-$\alpha$ island. This is followed by a transition to
region {\bf d} and then back to the $\beta$-strand region. 

Finally, the fourth trajectory that is also common in our data set is the one displayed in Figure 13d. It is similar with the trajectory
described in Figure 13c, except that now the residue that is located in $\tt L$-$\alpha$ island causes the transition from {\bf b} to {\bf d}
in Figure 12.

The Figures 12 and 13 reveal that the presence of a residue in the $\tt L$-$\alpha$ island relates to  a collective 
topology  of the backbone involving several residues. Since the definition of a  bond angle takes three $C_\alpha$ carbons 
and the definition of a torsion angle takes four  (see Figure 1), we conclude that the  topology of the trajectories in Figures 13a and 13b involve 
the interplay of seven residues while in 13c and 13d there are a total of eight residues present.

Finally, we have verified that all our results are independent of the data set we have used  by similarly
analyzing the proteins in the version v3.3 Library of chopped PDB files for representative  
CATH domains. The results are very similar. But in addition, we find that 
the  propensity is largest  in the (mainly-$\beta$) CA level classes 2.90, 2.160  where over 5$\%$ of all  residues are in the 
island. We also find that  any CA level family has  at least 1$\%$ of their residues in the island, except 1.40 where the 
single representative with PDB code 1PPR  has no residues in the island.  

\section{Conclusion}

We have investigated the non-glycyl  residues that are located in the $\tt L$-$\alpha$ region of the Ramachandran  map.
{\it Independently} of the amino acid,  we find that in the Discrete Frenet frames of the $C_\alpha$ carbons 
the corresponding side-chain  $C_\beta$ carbons are localized  
in the same direction.  This universality in the orientation
persists  when we  investigate the $C_\gamma$ and $C_\delta$ carbons, 
the side chain $O$ and $N$ atom  in the case of ASN and ASP. 
The results  suggest that instead of reflecting {\it only} a local interaction between
a given backbone unit and its residue,  the $\tt L$-$\alpha$ island  
is associated with  a largely residue independent  backbone conformation
that involves the collective  interplay between several consecutive residues.

 When we proceed to analyze the distribution of those backbone bond and torsion angles  that are associated with
the links that both precede and follow a residue that is located in the   $\tt L$-$\alpha$ island, we find that  independently of the residue
these angles display very similar patterns. Since the definition of a  bond angle takes three $C_\alpha$ carbons 
and the definition of a torsion angle takes four, this prompts us to propose that the geometrical structure
associated with the presence of a residue in the  $\tt L$-$\alpha$ island  reflects the 
interplay of at least seven  consecutive backbone units. In particular, we have not been able to pin-point any  
obvious local  reason (charged, polar, acidic, hydrophobic/philic) to explain the presence or absence of a residue   
on the $\tt L$-$\alpha$ region. 

Our approach is based on a novel method to depict proteins. In the course of our analysis we have been able to
observe several systematic patterns including anomalies in the PDB data, suggesting that the method we have utilized has a potential
of becoming a valuable tool for both experimental and theoretical protein structure analysis  and fold prediction.

\vskip 0.5cm 
\section*{Acknowledgement}
We thank S. Hu and J. \.Aqvist for discussions.


\vskip 0.5cm


\begin{thebibliography}{}

\bibitem[Ramachandran {\it et~al}., 1963]{rama} Ramachandran, G.N., Raakrishnan, C and Sasisekharan,  V. (1963),  Stereochemistry of polypeptide chain configurations  Journal of  Molecular Biolology {\bf 7}  95Ð99


\bibitem[Hovm\"oller {\it et~al}., 2002]{hov} Hovm\"oller,  S., Zhou, T.,  and Ohlson, T. (2002) Conformations of amino acids in proteins  Acta Crystallographica  {\bf D58}  768Ð776 

\bibitem[Deane {\it et~al}., 1999]{deane} Deane, C.M., Allen, F.H.,  Taylor, R.  and Blundell, T.L. (1999) Carbonyl-carbonyl interactions stabilize the partially allowed Ramachandran conformations of asparagine and aspartic acid,  Protein Engineering {\bf 12} 1025-1028 

\bibitem[Allen {\it et~al}., 1998]{allen} Allen, F.H., Baalham, C.A., Lommerse, J.O.M. and Raithby, P.R. (1998)
Carbonyl-carbonyl Interactions can be Competitive with Hydrogen Bonds
Acta Crystallographia  {\bf B54}  320-329

\bibitem[Chakrabarti {\it et.al.}.,  2001]{review} Chakrabarti, P. and Pal, D. (2001) The interrelationships of side-chain and main-chain conformations in proteins
Progress in Biophysics \& Molecular Biology {\bf 76} 1-102


\bibitem[Robinson {\it et~al}., 2001]{deami}  Robinson,  N.E. and Robinson, A.B. (2001) Deamidation of human proteins, 
Proceedings of National  Academy of Sciences  USA {\bf 98 } 12409Ð12414.

\bibitem[McCuddena {\it et~al}., 2006]{race}  McCuddena, C.R.  and Kraus, V.B.  (2006) Clinical Biochemistry  {\bf 39} 1112-1130


\bibitem[Koo {\it et~al}., 1999]{prion}  Koo, E.H.,  Lansbury, P.T. and Kelly, J.W. (1999)
Proceedings of National  Academy of Sciences USA {\bf 96} 9989-9990 


\bibitem[Robinson  {\it et~al}., 2004]{deami2}   Robinson, N.E.  and Robinson, A. B. (2004)  {\it Molecular Clocks Ð Deamidation of Asparaginyl and Glutaminyl Residues in Peptides and Proteins}
Althouse Press (London)


\bibitem[Hu {\it et~al}., 2011]{frenet} Hu, S.,  Lundgren, M. and Niemi,A.J. (2011)  e-print arXiv:1102.5658v1 [q-bio.BM]

\bibitem[Hanson, 2006]{hansonbook} Hanson,  A.J. (2006)  {\it Visualizing Quaternions}, Morgan Kaufmann Elsevier (London)

\bibitem[Kuipers, 1999]{kuipers} Kuipers, J.B. (1999)  {\it Quaternions and Rotation Sequences: a Primer with Applications to Orbits, Aerospace, and Virtual Reality}, Princeton University Press (Princeton) 

 \bibitem[Bishop,  1974]{bishop} Bishop, R.L. (1974)
There is more than one way to frame a curve
 Americal Mathematical Monthly {\bf 82} 246-251


\bibitem[Hartenberg {\it et~al}., 1964]{dh} Hartenberg R.S. and Denavit,  J. (1964) Kinematic synthesis of linkages (McGraw-Hill, New York, NY, 1964)

 \bibitem[Faddeev {\it et~al}., 1986]{lattice}  L. Faddeev, L and Takhtajan, L.  (1986) Hamiltonian methods in the theory of solitons Springer-Verlag (Berlin)
 
 \bibitem[Berman {\it et~al}., 2007]{pdb}  Berman, H.M., Henrick, K., Nakamura, H.,  J.L. Markley, J.L., (2007)  
The Worldwide Protein Data Bank (wwPDB): Ensuring a single, uniform archive of PDB data.  
{\it Nucleic Acids Research}  {\bf 35},   (Database issue) D301 

\bibitem[Orengo {\it et~al}., 1997]{cath} Orengo, C.A., Michie, A.D., Jones, S., Jones, D.T., Swindells, M.B., Thornton, J.M. (1997). CATH--a hierarchic classification of protein domain structures,
{\it Structure} {\bf 5},  1093Ð1108.

 
 \bibitem[Weichenberger {\it et~al}., 2007]{errors}  Weichenberger, C.X. and Sippl, M.J. (2007) NQ-Flipper: recognition and correction of erroneous asparagine and glutamine side-chain rotamers in protein structures, Nucleic Acids Research {\bf 35} (Web Server Issue) W403-406
 

\end{thebibliography}
\end{document}